# An ultra-massive fast-spinning white dwarf in a peculiar binary system

S. Mereghetti,[1*] A. Tiengo,[1] P. Esposito,[1,2] N. La Palombara,[1] G. L. Israel,[3] L. Stella[3]

[1]INAF – IASF Milano, via E. Bassini 15, 20133 Milano, Italy [2]INFN, Sezione di Pavia, via A. Bassi 6, 27100 Pavia, Italy [3]INAF – Osservatorio Astronomico di Roma, via Frascati 33, 00040 Monteporzio Catone, Italy

To whom correspondence should be addressed. E-mail: sandro@iasf-milano.inaf.it

**White dwarfs typically have masses in a narrow range centered at about 0.6 solar masses (Msun). Only a few ultra-massive white dwarfs (M>1.2 Msun) are known. Those in binary systems are of particular interest because a small amount of accreted mass could drive them above the Chandrasekhar limit, beyond which they become gravitationally unstable. Using data from the XMM-Newton satellite, we show that the X-ray pulsator RX J0648.0-4418 is a white dwarf with mass > 1.2 Msun, based only on dynamical measurements. This ultra-massive white dwarf in a post-common envelope binary with a hot subdwarf can reach the Chandrasekhar limit, and possibly explode as a Type Ia supernova, when its helium-rich companion will transfer mass at an increased rate through Roche lobe overflow.**

HD 49798 / RX J0648.0-4418 is the only known binary system composed of an early type subdwarf star (the stripped core of an intermediate mass star that lost its hydrogen outer layers during non-conservative mass transfer) and a pulsating X-ray source. Its optical/UV emission is dominated by the subdwarf HD 49798 which, being very bright (V-band magnitude of 8.3, spectral type sdO5.5), has been extensively studied. The orbital period ($P_{ORB}$ =1.5476637(32) days), projected semi-major axis ($A_C$ *sin i* = (2.506+/-0.014)x$10^{11}$ cm) and optical mass function ($f_{OPT} = 4\pi^2/(GP_{ORB}^2)(A_C \sin i)^3$ =0.263 +/- 0.004 Msun) have been accurately determined (*1, 2, 3*), but the nature of the companion star remained completely unknown until periodic X-ray pulsations at P=13.2 s were discovered in 1996 (*4*). This demonstrated that the companion star was either a neutron star or a white dwarf. However, because its X-ray luminosity is much smaller than that expected from a neutron star accreting in the stellar wind of HD 49798 (*5*), it has been concluded that RX J0648.0-4418 is a white dwarf (*6*). The process that led to the formation of this peculiar system is still poorly understood.

The presence of fast X-ray pulsations makes the system equivalent to a double-lined spectroscopic binary, for which all the parameters and in particular the masses of the two components can be derived. With this objective, we observed HD 49798 / RX J0648.0-4418 for 44 ks on May 10-11, 2008 with the Newton X-ray Multimirror Mission (XMM-Newton) satellite. The data were obtained with the EPIC instrument which provides imaging with high sensitivity and medium resolution spectroscopy in the 0.15-12 keV energy range (*7, 8*). Based on the optical ephemeris (*3*), the observation was scheduled so as to include the orbital phase at which the X-ray eclipse could be expected (Φ=0.75). Previous X-ray observations (*4, 9*) did not cover this orbital phase. Our X-ray light curve (Fig.1) shows the presence of an eclipse lasting ~1.3 hours, which allowed us to refine the measurement of the orbital period (see caption of Fig.1) and derive the system's inclination.

The X-ray pulsations from RX J0648.0-4418 are detected in the XMM-Newton data, with a broad single pulse and a pulsed fraction of 62% in the 0.15-0.4 keV energy range (fig.1, inset). By cross-correlating the folded light curves obtained at ten different orbital phases with an average template, we determined the delays in the times of arrival of the pulses induced by the orbital motion (fig. 1). By fitting them with a sinusoid (optical/UV spectroscopy indicates that the orbit is circular (*1, 3*)) we derived the projected semi-major axis $A_X \sin i$ = 9.78+/-0.06 light-s. This corresponds to an X-ray mass function $f_X = 4\pi^2/(GP^2_{ORB}) (A_X \sin i)^3$ = 0.419+/-0.008 Msun that, combined with the optical mass function, yields the masses of the two stars as a function of the orbital plane inclination (fig. 2). The inclination can be constrained by the duration of the X-ray eclipse, because it is related to the radius $R_C$ of HD 49798 by $(R_C/a)^2 = \cos^2 i + \sin^2 i \sin^2\Theta$ (a is the orbital separation and $\Theta$ is the eclipse half angle). The value $R_C$=1.45+/-0.25 solar radii (Rsun) derived for HD 49798 through optical spectroscopy (*2*) implies an inclination in the range 79-84 degrees (fig. 2). The corresponding masses are $M_X$=1.28+/-0.05 Msun for the white dwarf and $M_C$ = 1.50 +/- 0.05 Msun for HD 49798. Independently of the dimensions of HD 49798, a firm lower limit of 1.2 Msun (2$\sigma$ c.l.) on the white dwarf mass is obtained for the extreme case i= 90 degrees, making RX J0648.0-4418 one of the most massive white dwarfs currently known ( *10, 11, 12, 13, 14*).

This peculiar system is thought to be the outcome of a common envelope evolution. HD 49798 probably resulted from the evolution of a ~8-9 Msun progenitor that began to fill its Roche lobe before helium ignition and is currently in a core helium-burning phase ( *2, 15*). Alternatively, it could have a degenerate carbon-oxygen core and be in the shell helium-burning phase (*6*). The mass we derived for HD 49798 is consistent with both possibilities, and it rules out the case of a shell hydrogen-burning star resulting from a progenitor of mass <3.5 Msun that lost its hydrogen envelope while on the thermally pulsing asymptotic giant branch (*6*). The future evolution of HD 49798 will most likely lead to an increased mass transfer rate onto the white dwarf through Roche-lobe overflow during a semi-detached phase (*15, 16*). Theoretical evolutionary models of similar systems show that the transfer of helium-rich material can proceed at a rate of $10^{-6}$ –$10^{-5}$ Msun yr$^{-1}$, leading to an accreted mass of a few tenths of solar masses *(16)*. This would push the already massive white dwarf above the Chandrasekhar limit and possibly trigger a Type Ia supernova explosion.

Our analysis confirms that the X-ray spectrum of RX J0648.0-4418 is very soft (*4*), being well fitted by a blackbody with temperature kT=39+/-1 eV and an additional power law with photon index $\Gamma$~2, that accounts for only a small fraction of the flux (see the Supporting on-line material). The X-ray luminosity in the 0.2-10 keV energy range is ~2 $10^{31}$ (d/650 pc)$^2$ erg $s^{-1}$, where d is the source distance normalized to the value 650+/-100 pc derived from optical observations (*2*) and consistent with the parallax $\mu$=1.16+/-0.63 mas measured with Hipparcos (*17*). The wind properties of HD 49798 were derived previously by modelling of the P Cygni profiles of the UV lines, which indicates a mass loss Mdot in the $10^{-9}$ –$10^{-8}$ Msun $yr^{-1}$ range and a terminal wind velocity $V_{WIND}$ = 1350 km $s^{-1}$ (*5*), attained at a distance of only ~1.7 stellar radii from the subdwarf. Orbiting HD 49798 at a distance a=4 $10^{11}$ $(M_X+M_C)^{1/3}$ cm ~ 8 Rsun, the white dwarf accretes a fraction $\eta \sim (R_A/2a)^2$ ~0.0003 of Mdot, where $R_A$~2 $10^{10}$ (1350 km $s^{-1}/V_{WIND})^2$ cm is the accretion radius *(18)*.The expected luminosity is thus $L_X$= G $M_X \eta$ Mdot / $R_{WD}$ ~ $10^{31}$ (1350 km $s^{-1}$ / $V_{WIND})^4$ (Mdot / $10^{-9}$ Msun $yr^{-1}$) (3000 km / $R_{WD}$) erg $s^{-1}$ (*18*), in good agreement with the observed value and implying that the X-ray luminosity of RX J0648.0-4418 can be interpreted in terms of accretion from the wind of the hot subdwarf companion. Mass accretion down to the white dwarf surface requires the magnetospheric radius, defined by balancing the magnetic pressure and the ram pressure of the infalling material, to be smaller than the corotation radius $R_{COR} = (G M_X P^2/4\pi^2)^{1/3}$ = 9 $10^8$ cm. This translates into a magnetic dipole moment of ~< $10^{29}$ (Mdot / $10^{-8}$ Msun $yr^{-1})^{15/16}$ (1350 km $s^{-1}$ / $V_{WIND})^{15/4}$ Gauss $cm^3$ (*19*).

The high mass, fast spin, and possibly also low magnetic field (*20*) of RX J0648.0-4418 likely result from a previous evolutionary phase in which the accretion of mass and angular momentum took place at a much larger rate than currently observed. This system could thus be the white dwarf analog of low mass X-ray binaries in which weakly magnetic neutron stars are spun-up to short periods and shine as recycled millisecond radio pulsars when accretion onto them stops (*21*). A system probably undergoing such a transition from accretion-powered X-ray binary to millisecond radio pulsar has been reported very recently (*22*).

Most white dwarf masses are derived with indirect methods (*23*), such as surface gravity estimates obtained by spectral modeling, or the measurement of gravitational redshift (*24*). These methods, beside depending on models and assumptions, provide combined information on mass and radius: hence the resulting masses cannot be used to independently determine the mass-radius relation. Because our mass determination is based on a direct dynamical measurement, RX J0648.0-4418 can be used to constrain the white dwarf's equation of state. The derived lower limit on the mass, together with the 13 s spin period, implies an upper limit of 6000 km on the white dwarf's radius, based on simple rotational stability considerations (*25*). This limit is consistent with the classical predictions of zero-temperature stars (*26*), as well as recently derived mass-radius relations for massive oxygen-neon white dwarfs that predict a radius $R_{WD}$~3000 km for $M_X$=1.2 Msun (*27, 28*).

29. Based on observations obtained with XMM-Newton, an ESA science mission with instruments and contributions directly funded by ESA Member States and NASA. We thank E. Bozzo for interesting discussions. We acknowledge partial support from the Italian Space Agency (ASI/INAF contract I/088/06/0)

**Fig.1**. X-ray light curve of RX J0648.0-4418 in the 0.15-0.4 keV energy range obtained with the EPIC instrument on the XMM-Newton satellite. According to the optical ephemeris ( *3*) the eclipse centre was predicted to occur at MJD = 54597.169 +/- 0.023. The small difference with respect to the observed value (MJD = 54597.189+/-0.015) allowed us to determine the orbital period $P_{ORB}$= 1.5476666+/-0.0000022 days, which is consistent with, but more precise than, the previous value $P_{ORB}$=1.5476637+/-0.0000032 days (*3*). The points indicate the time delays of the X-ray pulsations measured at different orbital phases. They are induced by the orbital motion of the X-ray source and are well fitted by a sinusoidal curve with period fixed at $P_{ORB}$ and amplitude $A_X$ sin *i* = 9.78+/-0.06 light-s. **(Inset)** Pulse profile in the 0.15-0.4 keV energy range repeated for clarity over two cycles. The spin period P= 13.18425+/-0.00004 s has been determined with a standard phase fitting procedure.

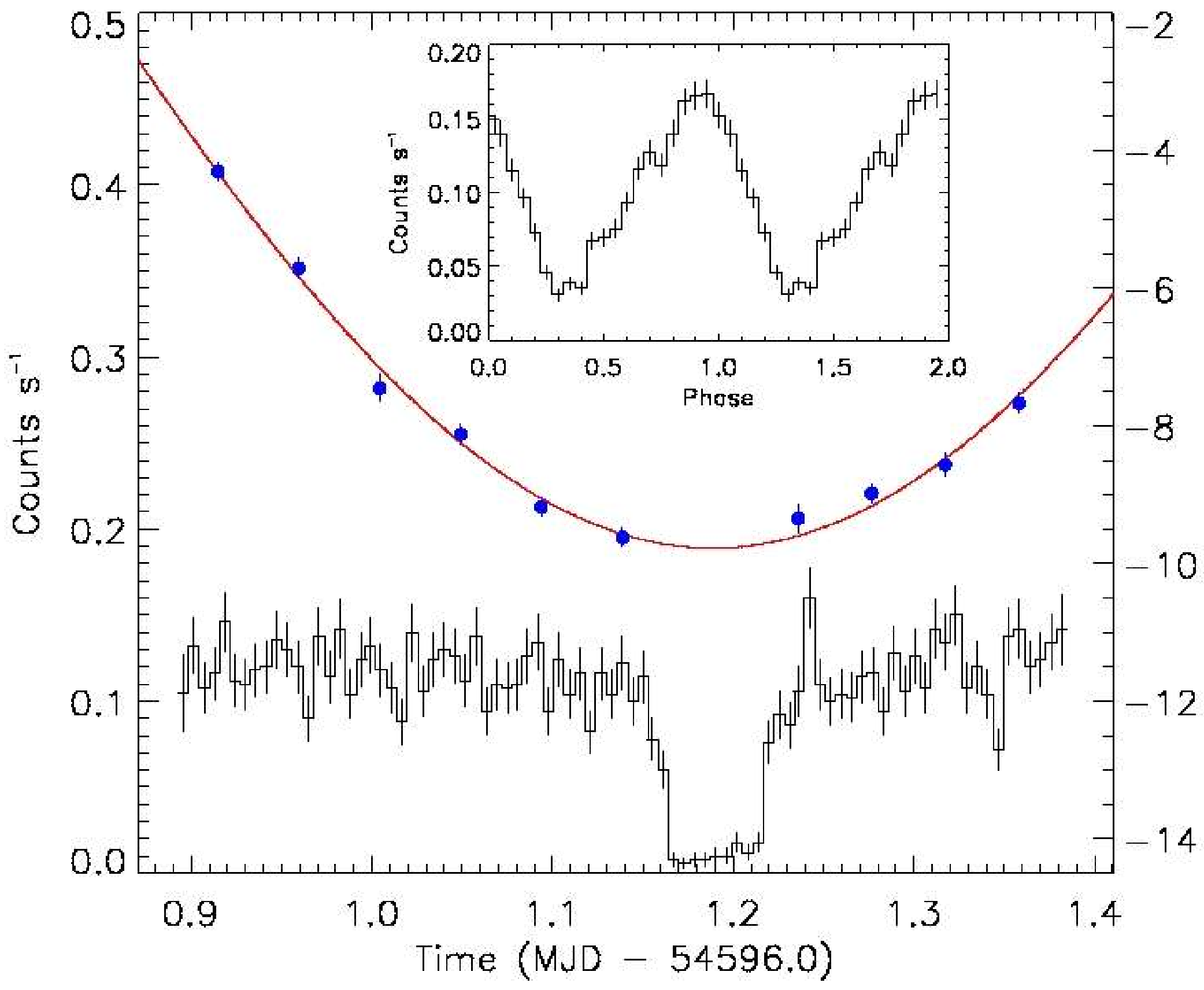

**Fig.2.** Masses of the HD 49798 subdwarf (right axis) and of its white dwarf companion (left axis) as a function of the orbital plane inclination. The solid line gives the values (and 1σ uncertainties, dash-dotted lines) corresponding to the optical mass function $f_{OPT} = M_X \sin^3 i /(1+q)^2 = 0.263 +/- 0.004$ Msun and the mass ratio $q = M_C / M_X = A_X \sin i / A_C \sin i = 1.17 +/- 0.01$. The dashed lines give the inclination angle derived from the duration of the X-ray eclipse for different values of the radius $R_C$ of HD 49798 (in units of Rsun). The distances (in kpc) corresponding to the different values of $R_C$ are also indicated.

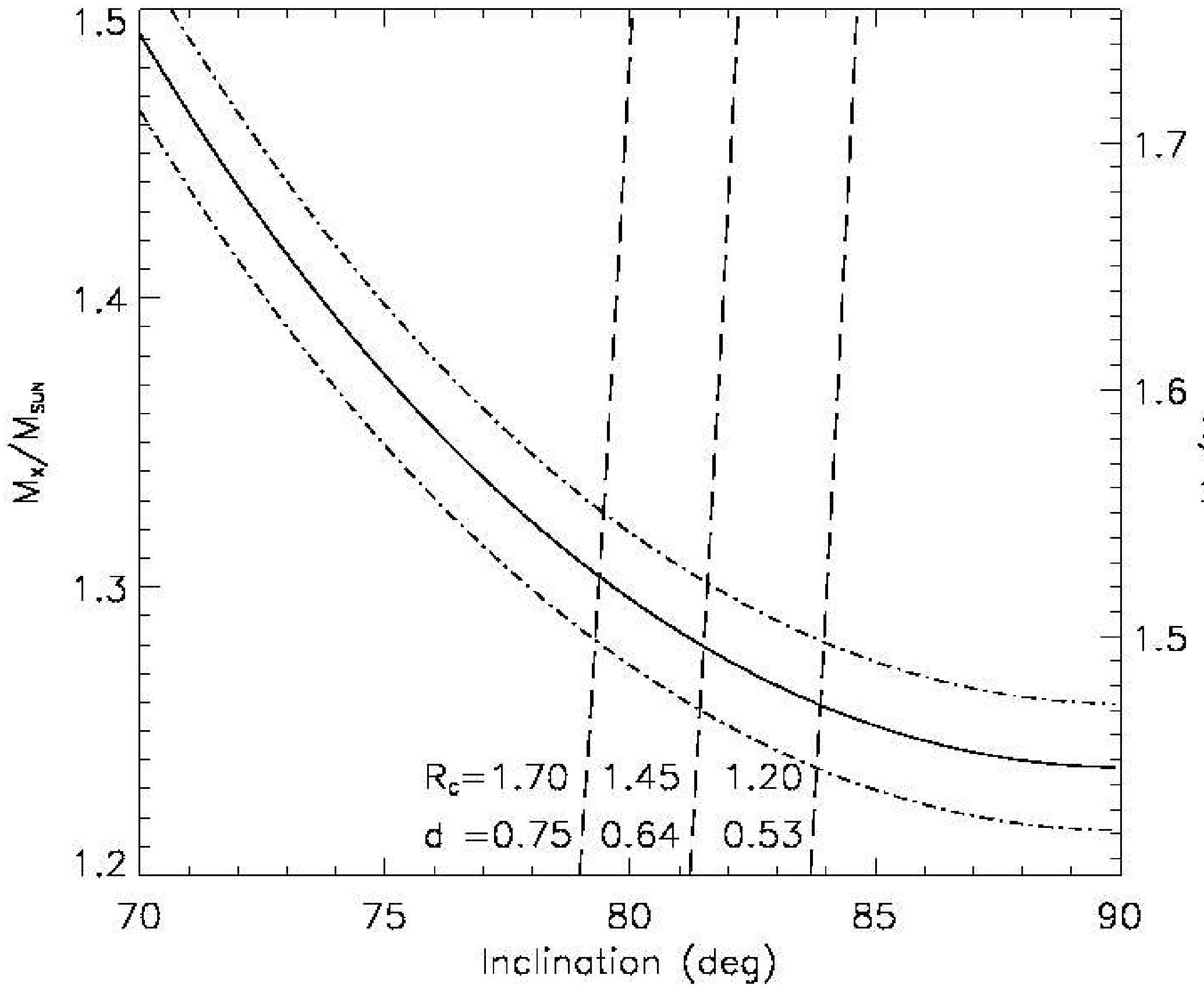

SUPPORTING ONLINE MATERIAL FOR

# An ultra-massive fast-spinning white dwarf in a peculiar binary system

S. Mereghetti,* A. Tiengo, P. Esposito, N. La Palombara, G. L. Israel, L. Stella

* To whom correspondence should be addressed. E-mail: sandro@iasf-milano.inaf.it

The observation of HD 49798 / RX J0648.0-4418 with the Newton X-ray Multi-mirror Mission (XMM-Newton) observatory started on 2008 May 10 at 21:07:45 UT and lasted 44.0 ks. The data used here were acquired with the European Photon Imaging Camera (EPIC) pn (*S1*) and MOS (*S2*) cameras, which were operated in `full frame' mode (with time resolution of 73.4-ms for pn and 2.6 s for the two MOS cameras) and mounted the medium thickness optical blocking filter. The data were processed with the XMM-Newton Science Analysis Software (SAS version 8.0). After correcting for dead time, we obtained a net exposure of 36.7 ks for the pn and 43.0 ks for the MOS cameras. The source photons for the spectral analysis were accumulated from a circular region (30 arcsec radius) centered on RX J0648.0-4418; a standard pattern selection was applied to filter the X-ray events (pattern 0-4 for pn and 0-12 for MOS). The background spectrum was extracted from a surrounding source-free region. A total of roughly 7,800 counts (~5,800 in the pn and ~2,000 in the sum of the two MOS) above the background were collected from the source in the 0.15-10 keV energy range. The eclipse time interval was excluded from the spectral analysis.

The X-ray spectrum of RX J0648.0-4418 is plotted in fig. S1. The spectral model ( $\chi^2$ = 180 for 159 degrees of freedom) describes the data as the sum of a blackbody curve (with temperature kT = 39.3+/- 1.3 eV and covering a surface with a 16 +/- 2 km radius, assuming a distance d=650 pc) and a power law (with photon index 1.96 +/- 0.03 and normalization (2.74 +/- 0.06) $10^{-5}$ photons $cm^{-2}$ $s^{-1}$ $keV^{-1}$ at 1 keV); the column density of neutral absorbing gas along the line of sight is $N_H$ = (1.7 +/- 0.6) $10^{19}$ $cm^{-2}$. All errors are at 1 σ. The corresponding observed flux in the 0.2-10 keV energy range is 4.7 $10^{-13}$ erg $s^{-1}$ $cm^{-2}$.

The high quality EPIC spectra rule out fits with a steep power law and/or lower temperature blackbody components that were consistent with previous data of lower statistical quality obtained with ROSAT and would imply a higher luminosity if extrapolated to the UV energy range (*S3*).

In principle, accretion onto a neutron star with mass 1.4 Msun and 10 km radius would produce a luminosity of 1.3 $10^{34}$ (1000 km $s^{-1}$ / $V_{WIND})^4$ (Mdot / $10^{-9}$ Msun $yr^{-1}$) erg $s^{-1}$ , clearly inconsistent with the observed value for any reasonable distance and wind parameters ($V_{WIND}$ ~1350 km $s^{-1}$ is the wind velocity at the neutron star distance and Mdot ~$10^{-9}$—$10^{-8}$ Msun $yr^{-1}$ is the mass loss rate from HD 49798 (*S4*)). However, a neutron star with magnetic dipole moment $\mu$ > 1.5 $10^{29}$ (Mdot / $10^{-9}$ Msun $yr^{-1})^{1/2}$ (1000 km $s^{-1}$ / $V_{WIND}$) G $cm^3$ would be in the supersonic propeller regime. Thus, for typical values of $\mu$, the accreting matter is halted at the magnetospheric boundary and it would be impossible to explain the large observed pulsed fraction (~60%). For a small range of magnetic moment values (down to $\mu$ > 0.3 $10^{29}$ (Mdot / $10^{-9}$ Msun $yr^{-1})^{15/16}$ (1000 km $s^{-1}$ / $V_{WIND})^{15/4}$ G $cm^3$ ), however the system would be in the sub-sonic propeller regime (*S5*), in which some accretion onto the neutron star is possible through the Kelvin-Helmholtz instability (*S6*). In this case we expect a luminosity within the range from ~1.5 $10^{32}$ ($\eta_{KH}$ /0.1) ($\varepsilon$/0.01)(Mdot / $10^{-9}$ Msun $yr^{-1}$) (1000 km $s^{-1}$ / $V_{WIND})^{10/7}$ erg $s^{-1}$ to ~4 $10^{34}$ ($\eta_{KH}$ /0.1) ($\varepsilon$/0.01) (Mdot / $10^{-9}$ Msun $yr^{-1})^{11/8}$ (1000 km $s^{-1}$ / $V_{WIND})^{11/2}$ erg $s^{-1}$ (*S6*) ($\eta_{KH}$ is the Kelvin-Helmholtz instability efficiency, and $\varepsilon$ is a function of the density increase across the magnetospheric boundary) . Also for the value $\varepsilon$~0.01, computed with conservative assumptions (*S6*), the predicted luminosity is much higher than the observed value. Furthermore, also in this case it would be difficult to account for the very large pulsed fraction.

We finally note, that in a less conventional alternative interpretation, the X-ray emission from RX J0648.0-4418 could be powered by the loss of rotational energy. In fact , with a rotation period of only 13.2 s, RX J0648.0-4418 is the fastest spinning white dwarf known. Based on a reanalysis of the X-ray observations obtained with ROSAT in 1992 (*S3*) and with XMM-Newton in 2002 (*S7*), taking into account the newly derived orbital parameters, we cannot exclude a spin period derivative dP/dt as large as 4 $10^{-13}$ s $s^{-1}$. For a moment of inertia $I_{WD} = 10^{50}$ g $cm^2$ the corresponding upper limit on the spin-down luminosity is $I_{WD}\ 4\pi^2\ P^{-3}$ dP/dt ~ 7 $10^{35}$ erg $s^{-1}$, which is much higher than the observed luminosity. While in this interpretation the power law component in the X-ray spectrum could be non-thermal rotation-powered emission, the highly pulsed, soft thermal-like emission component would be difficult to account for. We therefore favour the accretion scenario.

**Fig. S1**. EPIC/XMM-Newton spectrum of RX J0648.0-4418. **Top:** Data points and best-fitting spectral model. **Bottom:** residuals from the model in units of standard deviation.

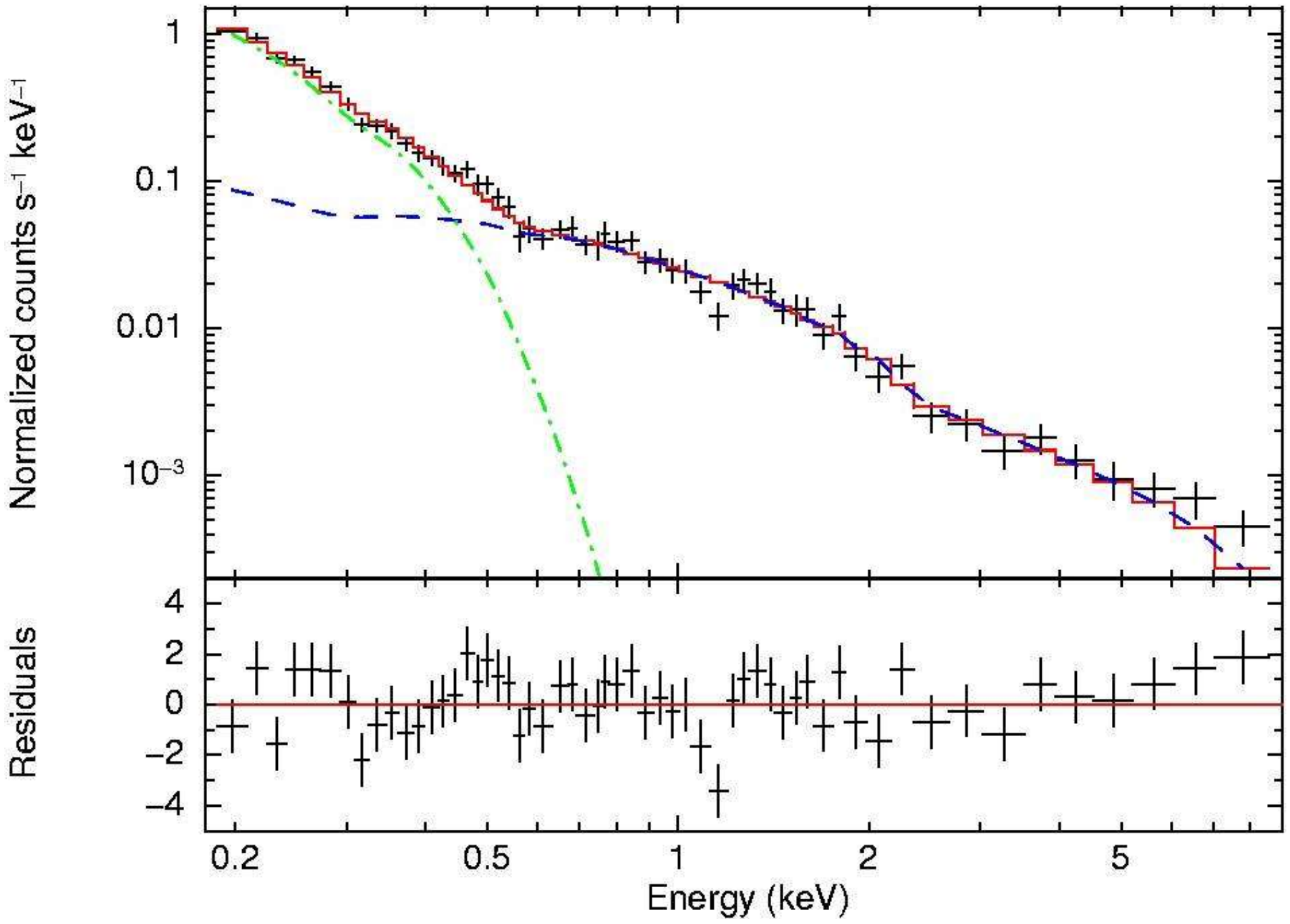